\documentclass[a4paper,fleqn,usenatbib]{mnras}

\usepackage{newtxtext,newtxmath}

\usepackage[T1]{fontenc}
\usepackage{ae,aecompl}

\usepackage{graphicx}	
\usepackage{amsmath}	
\usepackage{amssymb}	

\title[Bistability of exo-climates]{Climate bistability of Earth-like exoplanets}

\author[Murante et al.]{
Giuseppe Murante,$^{1,2}$\thanks{E-mail: Giuseppe.Murante@inaf.it}
Antonello Provenzale,$^{2,4}$
Giovanni Vladilo,$^{1}$
\newauthor
Giuliano Taffoni,$^{1}$
Laura Silva,$^{1}$ 
Elisa Palazzi,$^{3}$
Jost von Hardenberg,$^{5,3}$
\newauthor
Michele Maris,$^{1}$
Elisa Londero,$^{1}$
Cristina Knapic,$^{1}$
and Sonia Zorba$^{1}$
\\
$^{1}$INAF, Osservatorio Astronomico di Trieste, Via Tiepolo 11, 10134 Trieste (Italy) \\
$^{2}$ Institute of Geosciences and Earth Resources (IGG), National Research Council (CNR), Via Moruzzi 1, 56124 Pisa (Italy)\\
$^{3}$Institute of Atmospheric Sciences and Climate, National Research Council (ISAC-CNR), Torino (Italy)\\
$^{4}$ European Institute of Astrobiology (a virtual research infrastructure)\\
$^{5}$ Politecnico di Torino, Corso Duca degli Abruzzi 24, 10129 Torino (Italy)
}

\date{Accepted XXX. Received YYY; in original form ZZZ}

\pubyear{2019}

\begin{document}
\label{firstpage}
\pagerange{\pageref{firstpage}--\pageref{lastpage}}
\maketitle

\begin{abstract}
Before about 500 million years ago, most probably our planet
experienced temporary snowball conditions, with continental and sea
ices covering a large fraction of its surface. This points to a
potential bistability of Earth's climate, that can have at least two
different (statistical) equilibrium states for the same external
forcing (i.e., solar radiation). Here we explore the probability of
finding bistable climates in earth-like exoplanets, and consider the
properties of planetary climates obtained by varying the semi-major
orbital axis (thus, received stellar radiation), eccentricity and
obliquity, and atmospheric pressure. To this goal, we use the
Earth-like planet surface temperature model (ESTM), an extension of 1D
Energy Balance Models developed to provide a numerically efficient
climate estimator for parameter sensitivity studies and long climatic
simulations. After verifying that the ESTM is able to reproduce Earth
climate bistability, we identify the range of parameter space where
climate bistability is detected. An intriguing result of the present
work is that the planetary conditions that support climate bistability
are remarkably similar to those required for the sustainance of
complex, multicellular life on the planetary surface. The
interpretation of this result deserves further investigation, given
its relevance for the potential distribution of life in exoplanetary
systems.\\
\end{abstract}




\section{Introduction}
\label{sec:intro}

Rocky, earth-like exoplanets lying in the habitable zone of their host stars are
the most promising targets to search for life outside the Solar
System. Because of their small masses and sizes, only a small number
of rocky planets have been detected in the surveys carried out with
the radial-velocity \citep{Mayor11} and transit \citep{Batalha14}
methods so far. Nevertheless, planets with Earth-like masses and sizes
appear to be intrinsically more frequent than giant planets once the
statistics are corrected from the effects of observational bias
\citep{Mayor11, ForeMack14, Burke15}. Therefore we expect many rocky
planets to be discovered in forecoming observational surveys. As
statistics increases, the emphasis of the studies will be on the
characterization of the planetary properties and the selection of
targets with high probability of being habitable for further
analysis. Eventually, the selected planets will be the best targets
for follow-up observations aimed at looking for the presence of
atmospheric compounds of biological origin \citep{Kaltenegger15}.

In this general scenario, it is important to explore the impact of
planetary conditions that may affect the climate and habitability of
earth-like exoplanets. Two of such conditions are taken into account in the
classic definition of the habitable zone, namely the spectral energy
distribution of the host star and the planetary insolation, the latter
depending on the orbital semi-major axis and on stellar
luminosity. However, planetary climate and habitability are determined
by many other physical quantities, such as orbital eccentricity,
surface gravity acceleration, atmospheric pressure and composition,
obliquity of the rotation axis, rotation rate, and surface
distribution of oceans and continents \citep{Pierre10, KastingB}.
In this framework, a particularly interesting issue is the
possible existence of multiple stable solutions, for example in terms
of multiple stable equilibrium states such as those discussed for
Earth and found in numerical simulations of the terrestrial climate
\citep{Budyko69, Sellers69, Rose09, Ferreira11, Linsen15}.
If multiple stable states exist, then
the planet can end up in one or the other depending upon the initial
conditions. Similarly, external disturbances and the internal turbulent
dynamics of the climate system can push the planet out of one state
into another stable solution \citep{Benzi82}. For Earth, this mechanism has been
advocated for rationalizing the existence of snowball-like solutions
and warm states \citep{Zalia10}.  If multiple stable states
exist, then observing a planet that is temporarily in a snowball state
does not necessarily imply that the planet is not habitable.  Indeed,
some authors claimed that temporary snowball states are a requisite
for the evolution of metazoans with high energy demands \citep{RareEarth}.

In a modelling framework, many of the quantities determining planetary
climates are hard to measure with present-day observational facilities
and must be treated as free parameters. On top of the difficulty of
exploring a multi-parameter space of planetary conditions, one must
also face the general issue of how to model planetary climates, for
example, in terms of the level of complexity of the selected
models. Here, the choice ranges from extremely simplified
zero-dimensional Energy Balance Models (EBM), treating the planet as a
homogeneous point in cosmic space, see e.g. \cite{North81, Ghil94}, to 
fully-coupled ocean-atmosphere Global Climate Models (GCM), see e.g.
\cite{primer, trenberth}. 
Between EBMs and GCMs, an entire spectrum
of options is available, bearing in mind that the more complex the
model is, the larger is the amount of information needed to obtain
meaningful results (e.g., the position of the continents and their
orography). Given the paucity of information that is currently
available on the characteristics of exoplanets, in past years we
developed a climate model with an intermediate level of complexity,
namely the Earth-like planet Surface Temperature Model (ESTM)
discussed by \citet{Vladilo15} and succintly summarized in Section
3.

In the present work we use the ESTM to investigate what parameter
range is associated with the presence of bistability, and compare such
range with the conditions for habitability (using the liquid water
criterium). This analysis leads to some considerations on the
potential links between bistability and habitability for earth-like
planets. The presence of bistability is explored by determining the
final state (snowball or warm) depending on the initial temperature
conditions for a given set of planetary parameters. {The impact of the
  ice-albedo feedback in the context of bistability is specifically
  addressed.}

The systematic exploration of a broad space of planetary parameters,
as considered in our study, requires running a very large number of
climate simulations. The choice of models with an intermediate level
of complexity is thus the most suitable one, since these simple
climate models are characterized by a very low usage of CPU. In this
work, we explored approximately $10^5$ cases using the ESTM. The
current work extends and completes a previous study of climate
bistability in aquaplanets reported by \citet{Kilic17} using a
simplfied GCM.

The paper in structured as follows. In Section \ref{sec:bistabebm} we
introduce the general problem of bistability in the context of
simplified climate models while in Section \ref{sec:estm} we provide a
brief description of the ESTM. Section \ref{sec:dataset} presents the
dataset used to build up our collection of climate simulations. In
Section \ref{sec:earth} we discuss the bistability of Earth climate,
as simulated by the ESTM, and in Section \ref{sec:prob} we explore the
range of parameter values where climate bistability in earth-like
exoplanets is potentially present, comparing it with the range where
habitability is possible. Finally, Section \ref{sec:conclu} provides a
discussion of the results and some conclusions.

\section{Bistability in Energy-Balance Models}
\label{sec:bistabebm}

Energy Balance Models (EBMs) are simplified thermodynamical
descriptions of the surface Earth temperature \citep{Budyko69, Sellers69}, 
and they can be considered as the simplest global climate
models \citep{North81, Saltzman02}. The main point of EBMs is
that the planetary surface temperature is obtained from a
suitably-written form of the first principle of thermodynamics, where
the temporal variation of the temperature $T$ is given by the
difference between the absorbed solar radiation (for Earth, mostly in
the visible part of the spectrum) and the (infrared) radiation emitted
by the planet. The absorbed radiation is the fraction of received
stellar radiation that is not reflected back to space (that is, the
fraction $1-A$ where $0 \le A \le 1$ is the planetary albedo). At
(statistical) equilibrium, the absorbed and emitted radiation at
global level are balanced.

{EBMs can be zero- or one-dimensional}. Simplest versions are
zero-dimensional, that is, they describe the globally averaged planet
surface temperature, which depends only on time $t$. More refined
versions are spatially one-dimensional, that is, they describe the
surface temperature as a function of latitude and time. In this
version, at each latitude, besides absorbed and emitted radiation, one
must include also the divergence of the meridional heat flux, {which
  accounts for the latitudinal heat transport}. Often, the heat flux
is parameterized by the gradient of temperature, which results in a
heat diffusion term in the temperature equation 
\citep{North81,Adams, Wood}. In such conditions,
most of the uncertainty depends on the proper parameterization of the
diffusion coefficient, which in principle can depend on latitude,
temperature, and time. In a recent paper \citep{Vladilo15}, an
extended version of the simplest one-dimensional EBM was proposed and
tested. The current version of the model, called Earth-like planet
Surface Temperature Model (ESTM, {see Section \ref{sec:estm} for
  details}), uses a more refined definition of the diffusion
coefficient mimicking the effects of atmospheric circulation, a
latitudinal dependence of the fraction of continental areas, a more
precise parameterization of albedo from the ocean and the continents,
and a set of one-dimensional vertical column radiative-convective
calculations that define the emitted radiation at the top of the
atmosphere as a function of surface temperature and atmospheric
composition. Simplified prognostic equations for ice-sheet dynamics
and for sea ice are also included, as well as a simple prescription
for clouds.

In the absence of nonlinear feedbacks, EBMs usually provide a unique
(globally stable) equilibrium solution. That is, given the stellar
input, the planetary albedo, and the characteristics of the surface
and of the atmosphere, the equilibrium temperature where absorbed and
emitted radiation balance is uniquely defined.

Classic terrestrial EBMs, however, have long been used to provide a
simplified description of the possibility of multiple equilibrium
states for the same set of forcing conditions, such as the
glacial-interglacial conditions in the last three million years or the
emergence of Snowball Earth (or icehouse) states. To clarify this
point, we recall that multiple equilibria indicate the possibility of
different equilibrium solutions for the same values of external
forcing and internal system parameters. The system will then
dynamically settle on one specific equilibrium state depending on the
initial conditions. External disturbances or the internal natural
climate variability (not described in such simple EBMs) could push the
planet's climate from one state to the other.

For multiple equilibria to occur, a nonlinear feedback between
temperature and either reflected or emitted radiation (besides the
Stefan-Boltzmann law) must be introduced. The simplest and more
standard approach is to introduce a feedback between temperature and
albedo, based on the presence of ice cover \citep[e.g.][]{Ghil94}. Ice
albedo is much larger than that of the bare (or vegetated)
ground. Thus, for low surface temperature, the ice cover is more
extended than for higher temperatures, and the albedo at low T is
correspondingly higher. A typical parameterization of this effect is a
constant and high albedo below a critical temperature threshold (say a
few degrees below the freezing point), when the ice cover is
extensive, a constant and low albedo above a upper temperature
threshold representing the full melting of ice cover, and a linear
dependence between the two threshold. Alternatively, a hyperbolic
tangent dependence of albedo on temperature can provide similar
results.  With such albedo parameterization, for intermediate
insolation conditions EBMs may provide two stable equilibria
corresponding respectively to fully-glaciated (snowball) condition and
to an (almost) ice-free world, and an unstable equilibrium in between
these two extremes.

An important point concerns the role of stabilizing (negative)
and destabilizing (positive) feedbacks. Negative feedbacks tend to
damp the effects of external perturbations, leading to an equilibrium
point which can survive external disturbances. A classic example in
planetary climates is the homeostatic behavior observed in simple
models such as Daisyworld \citep{Watson83, Lenton01, Lenton02, Wood}. 
Here, even a change in the stellar output {(such as its increase with time)} 
can be balanced by a change in the surface albedo associated with the
dominance of different types of vegetation. By contrast, positive
feedbacks tend to amplify the effects of {initial} perturbations and
typically lead to multiple equilibria, as in the examples discussed
above, or to runaway situations as in the moist runaway greenhouse
effect.

More generally, the stability and bifurcations properties of EBMs and
Daisyworld-like models, also including their suitability for the Earth's
climate, have been widely explored, see e.g. \cite{Ghil1976, Rombouts2015}. 
The parameter space for climate stability of Earth-like planets has also been 
investigated \citep{Alberti2018}. In addition, the evidence of striped patterns 
in one-dimensional models could be seen as a manifestation of
spatial multiple solutions \citep{Adams}, closely related to the
existence of positive and negative feedbacks changing stability properties
\citep{Alberti2015}.

The existence of multiple climate states has been explored also with
more detailed Global Climate Models
\citep{Rose09, Ferreira11, Linsen15}, confirming end
extending the results obtained with simpler models. More recently, \citet{Kilic17}
have used an Earth System model of Intermediate
Complexity \citep[the ``Planet Simulator'',][]{Lunkeit11} to study the
presence of multiple stable states in aquaplanets, for different
values of stellar irradiation and obliquity, finding, also in this
case, the potential coexistence of warm and snowball states.

Before closing, a few comments are in order. First, ice-temperature
feedback is not the only one that can produce multiple equilibria. 
Indeed, there are a lot of feedbacks to be considered in building up a
climate model that could lead to changes in the stability conditions
and to multiple steady state solutions, see e.g. \cite{Ghil2019}. 

For example, vegetation usually is darker than sandy soils. From this
observation, \citet{Charney75} proposed the possibility of multiple
equilibria in arid regions, associated with the presence or absence of
vegetation. Extending this approach, \citet{DAndrea06} and
\citet{Baudena08} showed that multiple equilibria in arid continental
regions can also emerge from the fact that the presence of vegetation
leads to larger evapotranspiration (and thus more unstable atmospheric
stratification) than bare soil. This approach was later extended to a
simplified zero-dimensional, columnar model representing a whole arid
planet, showing that the presence of vegetation could sustain a
hydrological cycle, which would instead disappear for bare ground
\citep{Cresto13}. Finally, also the albedo effect of cloud
cover has been shown to be able to increase the number of coexisting
multiple equilibria.

A second point concerns the possibility of oscillating or chaotic
solutions, something well known in low-order dynamical systems but not
very common in EBMs. Early attempts to obtain self-sustained
oscillatory solutions in EBMs considered coupling ice sheets,
atmosphere and ocean \citep{Kallen79} or an EBM coupled with
ice-sheet dynamics and the slow isostatic response of the viscous flow
in the upper mantle \citep{Ghil81}. Also for limit cycles
(self-sustained oscillations) and chaotic solutions, it is possible to
have multiple solutions for the same external forcing levels, as
discussed by \citet{Lorenz90}. In general, however, obtaining
self-sustained oscillations requires at least two coupled dynamical
equations, and it is a behavior that is not expected in the simplest
EBM or ESTM approaches.

\section{THE ESTM MODEL}
\label{sec:estm}

The model adopted here is the Earth-like planet surface temperature
model (ESTM) described in detail by \citet{Vladilo15}. The core of
the model is a one-dimensional Energy Balance Model (EBM) fed by
multi-parameter physical quantities and written for an elliptical
orbit and taking into account the planet axial tilt (thus,
incorporating the seasonality of the stellar radiation intercepted by
the planet). The dynamical variable is the planet surface temperature
$T(\phi,t)$, that is a function of latitude $\phi$ and time $t$. All
relevant quantities are thus zonally averaged. As an improvement over
classic EBMs, the vertical transport in ESTM is described using
single-column atmospheric calculations, while meridional transport is
parameterized based on the results of full 3D climate model
simulations.

By assuming that the heating and cooling rates and the meridional heat
transport are balanced at each latitude, one obtains an energy balance
equation that is used to calculate $T(\phi,t)$. The most common form
of EBM equation \citep{North81, WK97, Spiegel08, Pierre10, Gilmore14}

\begin{equation}
\label{eq:ESTM}
C\frac{\partial T}{\partial t}=S(1-A)-I+\frac{\partial}{\partial x}\left[D(1-x^2)\frac{\partial T}{\partial x}\right]
\end{equation}

where $x = \sin \phi$ and all terms are normalized per unit area. The
term on the left hand side of this equation represents the zonal heat
storage and describes the temporal evolution of each zone; $C$ is the
zonal heat capacity per unit area (North et al. 1981). The first term
on the right hand side representes the absorbed stellar radiation, $S$
is the incoming stellar radiation in $W/m^2$ and $A$ is the planetary
albedo at the top of the atmosphere, function of both temperature and
latitude.

In the ESTM, the dependence of albedo $A$ on temperature is not
directly imposed, but it is naturally parameterized by the latitudinal
dependence of the ice cover on temperature. The total albedo at each
latitude is then given by the weighted average between the albedo of
the land or ocean and the (larger) albedo of the (continental or sea)
ice. This approach provides similar results {as} a direct
parameterization of albedo on temperature, but it has the advantage of
introducing a more physically-based description of the processes at
work. Thus, as it will be shown below, also for one-dimensional ESTM
there is the possibility of multiple equilibria whose existence relies
on a temperature-albedo feedback mechanism.

The second term on the right hand side of eq. (\ref{eq:ESTM}), $I$,
represents the thermal radiation emitted by each zone, also called
Outgoing Longwave Radiation (OLR). While in simple EBMs this is given
by the Stefan-Boltzmann law corrected for greenhouse effects, here the
OLR is obtained from detailed vertical column calculations taking into
account radiative-convective processes. The third term represents the
amount of heat per unit time and unit area leaving each zone along the
meridional direction \citep{North81}. It is called the
diffusion term because the coefficient $D$ is defined on the basis of
the analogy with heat diffusion, i.e.

\begin{equation}
\Phi\equiv-D\frac{\partial T}{\partial \phi}
\label{eq:heat}
\end{equation}

where $2\pi R^2\Phi\cos\phi$ is the net rate of energy transport
across a circle of constant latitude and $R$ is the planet radius
\citep[see ][]{Pierre10}.

Thanks to a physically-based parameterization, the ESTM features a
dependence on surface pressure, $p$, gravitational acceleration, $g$,
planet radius, $R$, rotation rate, $\Omega$, surface albedo, $a_s$,
stellar zenith distance, $Z$, atmospheric chemical composition, and
mean radiative properties of the clouds. In a typical simulation, the
ESTM generates the temporal dynamics of the surface temperature
$T(\phi; t)$ starting from a given initial conditions
$T_0(\phi)$. Full details on the ESTM are given in \citet{Vladilo15}.

In the ESTM different temperature-based habitability criteria can be
adopted. In this work, habitability is defined based on a liquid water
criterion for the presence of complex life \citep{Laura16}: that
is, potentially habitable conditions correspond to a temperature range
between the freezing point of water (for a given atmospheric pressure)
and an upper limit $T\approx 50$ C needed for the persistence of
active metabolism and reproduction of multicellular poikilotherms
(organisms whose body temperature and functioning of all vital
processes is directly affected by the ambient temperature). We thus
adopt a habitability index for complex life, $h_{050}$, representing
the mean orbital fraction of planetary surface that satisfies the
above temperature limits, see \citet{Laura16} for further details.

\begin{figure*}
\centering
\includegraphics[width=0.60\textwidth]{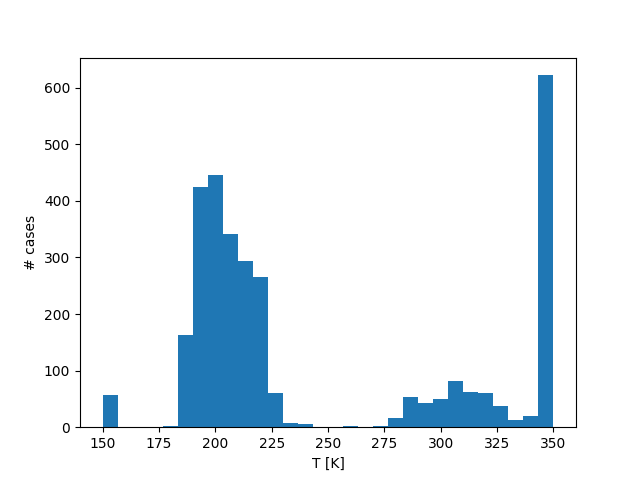}
\caption{Histogram of the final annual global average temperatures for all of our 86422 valid runs (including those that did not converge, but ended up in a runaway greenhouse situation or an automatic snowball one). Note that, clearly, the histogram is obtained using the number of {\it cases} not that of {\it quadruplets}. }
\label{fig:histT-all}
\end{figure*}

\begin{figure*}
\centering
\includegraphics[width=0.44\textwidth]{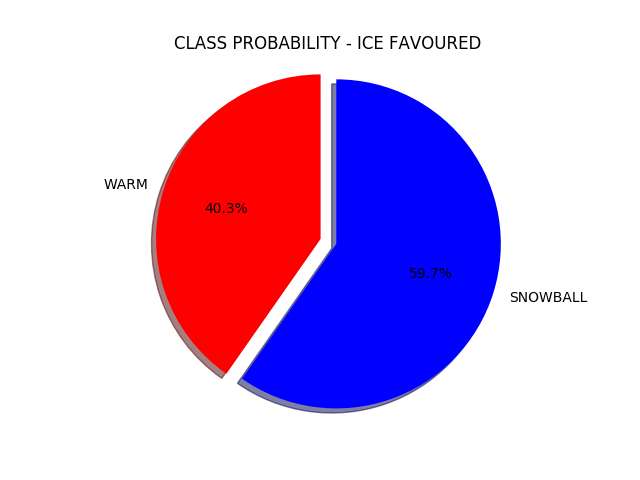}
\includegraphics[width=0.44\textwidth]{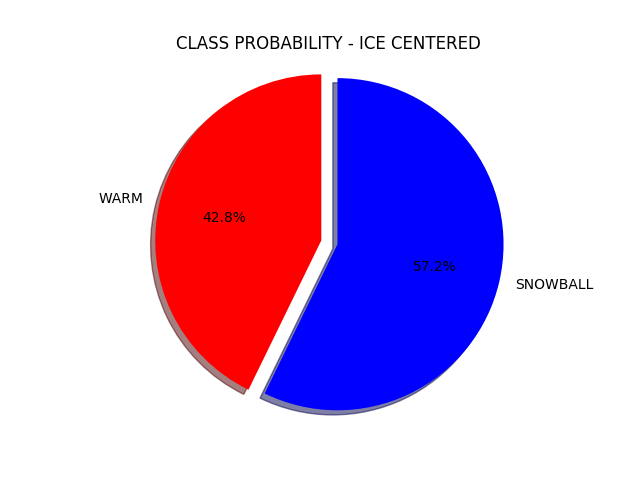}
\includegraphics[width=0.44\textwidth]{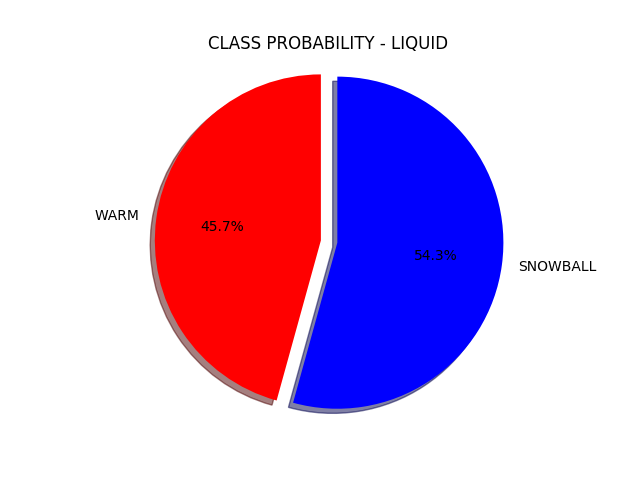}
\caption{Probability of having WARM or SNOWBALL cases, for the three different probability support defined in Section \ref{sec:dataset}. We integrated over the whole parameter set and did not distinguish stable from bistable cases.}
\label{fig:generalpies}
\end{figure*}

\begin{figure*}
\centering
\includegraphics[width=0.44\textwidth]{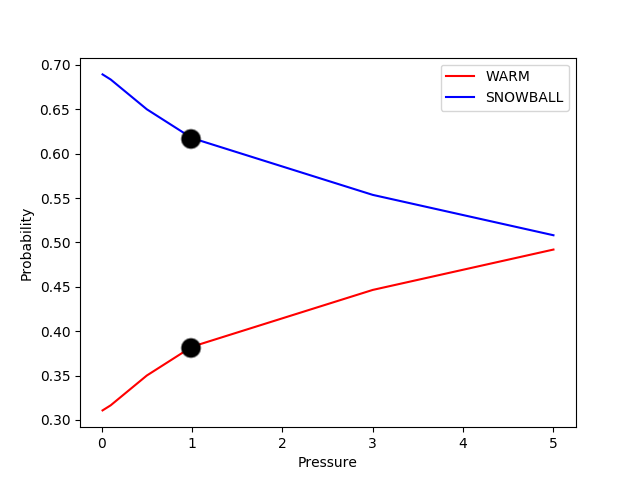}
\includegraphics[width=0.44\textwidth]{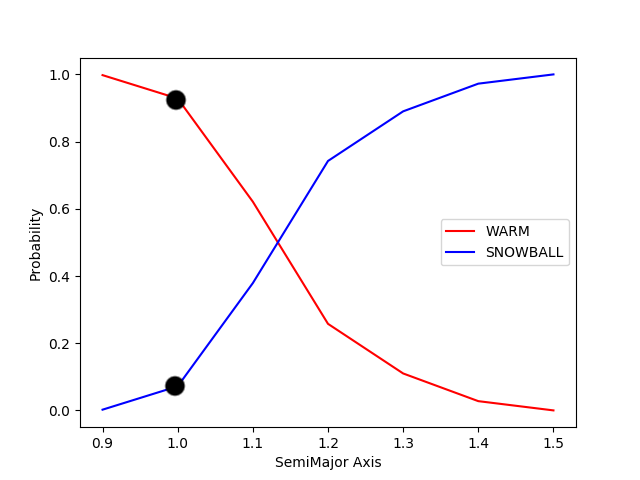}
\includegraphics[width=0.44\textwidth]{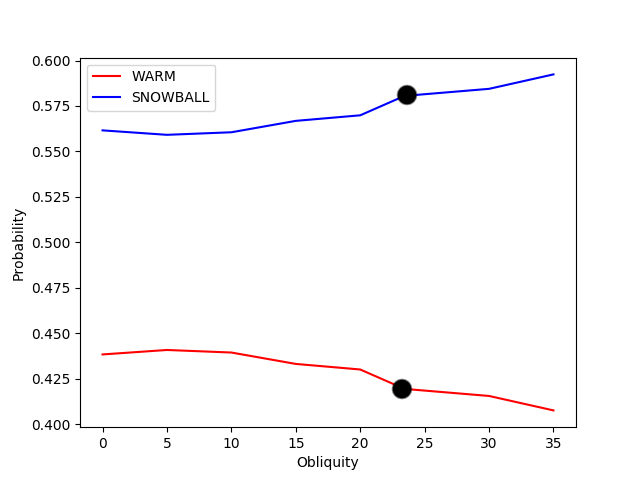}
\includegraphics[width=0.44\textwidth]{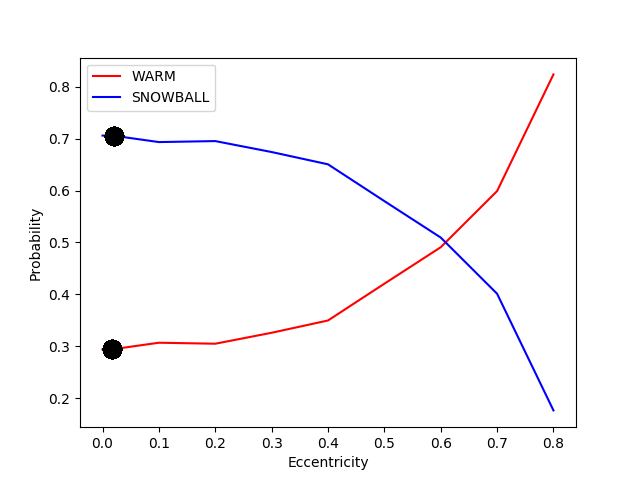}
\caption{Probability of having a WARM and SNOWBALL cases, as a function of our four parameters separately. For each  parameter, we calculated the probability using Equation \ref{sec:prob}, and estending the summations in Equation \ref{eq:dV} to all the indexes of the other four parameters. Upper-left panel: probability as a function of pressure; Upper-right panel: probability as a function of semi-major axis; lower-left panel: probability as a function of 
obliquity; lower-right panel: probability as a function of eccentricity. In red  the probability to end up in the WARM class, in blue that to end up in the SNOWBALL one. Here we show results for the {\it IceCentered} probability support. Black dots show the position of Earth in each plot.
}
\label{fig:plotParamIce}
\end{figure*}

\begin{figure*}
\centering
\includegraphics[width=0.77\textwidth]{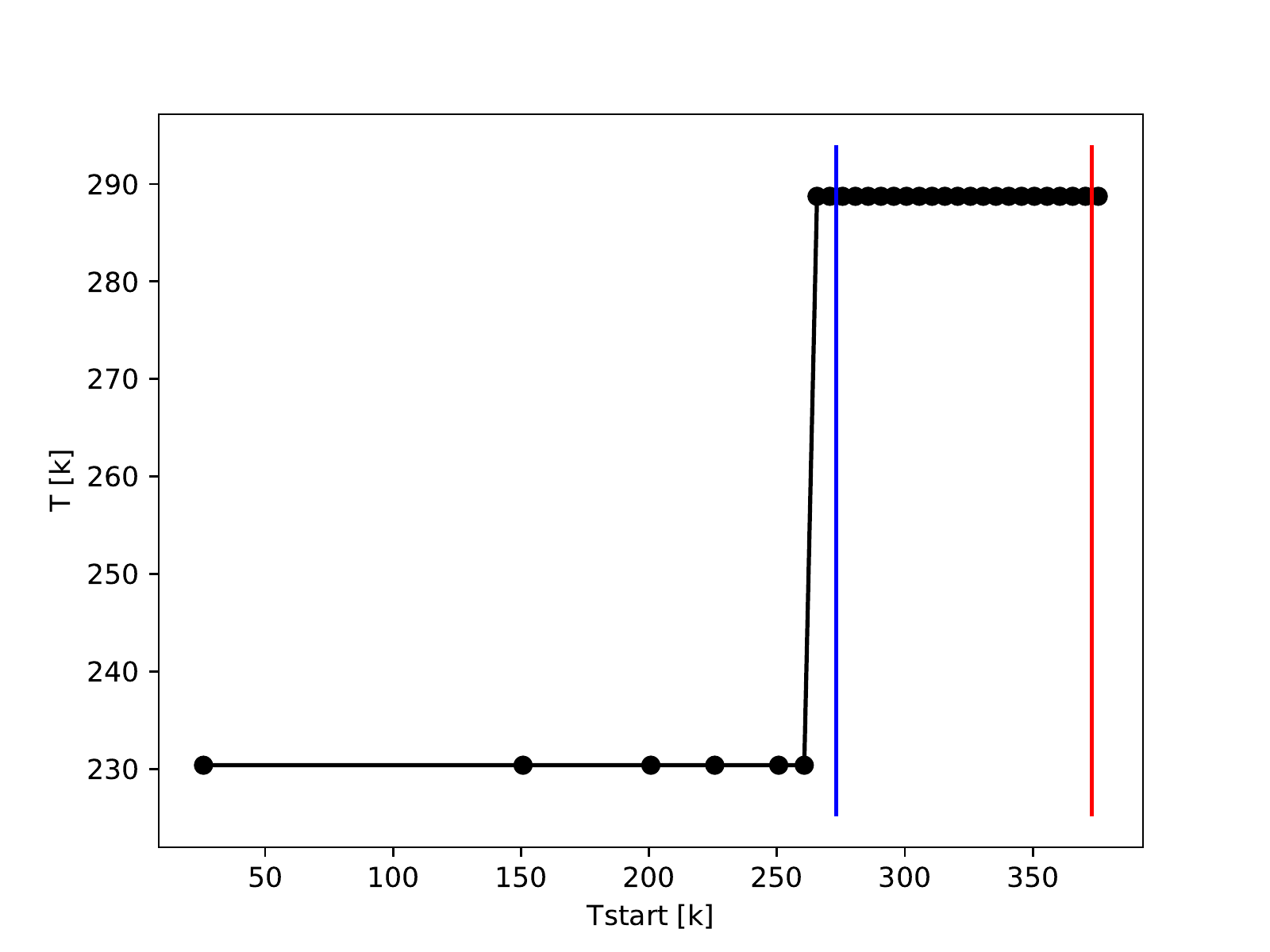}
\caption{Annual global average temperature in the stationary state, as a function of initial temperature for the Earth case, quadruplet with \{pressure, semi-major axis, obliquity, eccentricity\}=\{1.0, 1.0, 23.4393,0.01671022\}. The blue and red vertical lines mark the freezing and boiling temperature.}
\label{fig:earth}
\end{figure*}

\begin{table*}
	\centering
	\caption{ Parameters used for our ESTM model runs. Column 1: parameter; Column 2: values. Pressures in bar; Semi-Major axis in Astronomical Units; Obliquities in degrees; for Initial Iemperatures we give the index $i_T$ of $T_{\rm init} = T_{\rm freeze} + i_T \cdot dT$, see text for mode  details. Column 3: indexes used in Equation \ref{eq:dV}}
	\label{table:dataset}
	\begin{tabular}{lcl} 
		\hline
		Parameters & Value & index \\
		\hline
		Pressures & 0.01, 0.1, 0.5, 1.0, 3.0, 5.0 & $i_p$\\
        Semi-Major Axis & 0.9, 1.0, 1.1, 1.2, 1.3, 1.4, 1.5 & $i_r$  \\
		Obliquities & 0., 5., 10., 15., 20., 23.4393, 30., 35., & $i_o$\\
		Eccentricities &  0.0, 0.01671022, 0.1, 0.2, 0.3, 0.4, 0.5, 0.6, 0.7, 0.8 & $i_e$\\
		Initial temperatures &-50, -25, -15, -10, -5, -3, -2, -1,0,1,2,3,4,5,6,7,8,9,10,11,12,13,14,15,16,17,18,19,20 & $i_T$\\
		\hline
	\end{tabular}
\end{table*}

\section{The Dataset}
\label{sec:dataset}

We run the  ESTM varying four parameters, namely pressure, orbital eccentricity, inclination of the rotation axis and semi-major axis (SMA) of the orbit, for a total of $N_{\rm param}*N_{T_{\rm init}}=97440$ cases. $N_{\rm param}=3360$ is the number of independent simulations, each of which is run using $N_{T_{\rm init}}=29$ different initial temperatures. 

We set the initial temperatures using:
\begin{equation}
\label{eq:tinit}
T_{\rm init} = T_{\rm freeze} + i_T * dT
\end{equation}
where $T_{\rm freeze}$ is the pressure-dependent freezing point of water, and

\begin{equation}
\label{eq:dt}
dT = (T_{\rm boil} - T_{\rm freeze})/20
\end{equation}

here $T_{\rm boil}$ is the pressure-dependent water boiling
temperature and $i_T = 1, ..., 20$. Thus we have 20 different initial
temperatures in the pressure-dependent liquid water temperature
ranges. We added 9 more values at or below the freezing temperature of
water, since we determined that the onset of bistability in our model
can happen below such a temperature {(see Section
  \ref{sec:earth})}. In Table \ref{table:dataset} we list the
parameter values used.

All other model parameters, and in particular: radius of the planet,
mass of the planet, chemical composition of the atmosphere, rotation
period of the planet, type of star of the system, partial methane
(CH$_4$) and carbon dioxide (CO$_2$) pressures, are set to Earth
values. We used a fixed constant fractional ocean coverage of 70\%.

Given a set of parameters, usually the ESTM converges to an annual
oscillation, where the surface temperatures in each latitudinal band
and at each given position in the orbit have the same value for the
different years. We verified this behavior by asking that the global
annual average temperature (averaged over the latitudes at a given
position in the orbit) does not change by more than 0.01 K from one
year to the next. After convergence has been reached, we record a
number of model outputs; we focus our analyses on the global annual
surface temperature $T_{\rm fin}$ and the fractional ice coverage
$f_{\rm ice}$.

However, the model can exit for reasons different from convergence. If
more than 50\% of the planet surface has a temperature above the water
boiling point, we consider such a planet to have high probability of
entering a runaway greenhouse state and stop the numerical
simulation. In this case we assign a conventional final temperature of
350K. When the maximum temperature of the planet, considering all the
latitudinal bands and 48 position on the orbit, is lower than 200 K,
we also interrupt the run and consider the planet prone to reach a
permanent snowball state. In this case, the conventional final
temperature is set to 150 K; we called these cases ``automatic
snowballs'' and we implemented a forced exit in these cases, in order
to speed-up the calculations.

Moreover, there are a number of cases in which a run can end
traumatically. This happens: (i) if the total (dry plus water vapour)
pressure exceeds 10 bar; (ii) if the integration step becomes smaller
than 1 hour. Both cases correspond to situations where the hypotheses
that our model is based upon are not valid any more. Usually, these
traumatic exits happen for extreme values of some parameters, for
instance, for very high orbital eccentricities and/or small semi-major
axis. These runs must be discarded. Our dataset includes 289
quadruplets of parameters\footnote{We here define a quadruplet as the
  set of 4 parameters for which we run the 29 different numerical
  experiments corresponding to 29 different initial temperatures.}
leading to the kind of exits describe above.

{In a very limited number of cases, it may also happen that the run
  does not converge after a given number of orbits. We set this number
  to 100. The reason is that our continental ice model is based on the
  annual average temperature in each latitudinal band (owing to the
  slow response time of ice sheets) and in some situation the
  ice/albedo feedback may cause the albedo to have a non-periodic,
  non-converging behavior. This happened for 20 quadruplets, which we
  also discarded.}

At high pressures, it may happen that the initial temperature
corresponding to $i_T=-50$ in Equation \ref{eq:tinit} would be
negative (in Kelvin). This happened for 1125 quadruplets and we of
course discarded such initial conditions.

Finally, for 61 quadruplets, {\it some} of our runs with different
initial temperatures ended with a traumatic exit. We kept these cases
and considered only the runs that converged.

All together, we have a total of { 3051 complete or partial
  quadruplets} of parameters, for a total of { 86422 valid
  runs}.\footnote{We also realized a set of simulations with similar
  initial parameters, but always starting from an initial temperature
  of 275K ("cold start", see e.g. Spiegel 2009). In this set, in
  addition to the parameters explored here, we also varied the
  fractional ocean cover and the CO$_2$ partial pressure. Using this
  second set, we generated a public database. The database name is
  ARTECS (ARchive of TErrestrial-type Climate Simulations).  The
  database is reachable at the WEB address:
  http://wwwuser.oats.inaf.it/exobio/climates. Note that we are not
  using the full database in the present work; rather, this work can
  be seen as a preliminary step for using our full ARTECS database to
  study in details the habitability of earth-like planets.}

The final average temperatures for all the considered quadruplets
starting from the different temperatures used for initial conditions,
for a total of 86422 valid runs including those that did not converge
but ended up in a runaway greenhouse situation or an automatic
snowball one, are shown in figure \ref{fig:histT-all}. The figure
shows that there are two different classes of final average
temperatures, that can be neatly separated using a temperature
threshold of $T_{\rm thresh}=250$K. We thus define two classes of
solutions and call {\it WARM} those solutions having a temperature
higher than $T_{\rm thres}$, and {\it SNOWBALL} the other ones. Our
criterion for a run to be classified as SNOWBALL is that its final
global annual average temperature is lower than the water freezing
temperature $T_{\rm freeze}$, and the surface fraction covered by ice
must is grater than 90\%, otherwise its class is WARM. This is for
taking into account cases where an equatorial belt with temperatures
larger than the $T_{\rm freeze}$ coexist with large polar ice caps, so
that the {\it global} annual average temperature can be lower than
$T_{\rm freeze}$\footnote{Note that such cases could be considered as
  a different class, sometimes called {\it WATERBELT} in the
  literature, see e.g. \citet{Wolf17}). We will take into account such further
  classification in a forthcoming paper, dedicated to the explicit
  study of habitability}. We note that this never happens when the
final temperature is lower than 250 K.
 
Having defined the two classes of solutions, we need a definition for
the probability to end up in one of those classes. Our
four-dimensional parameter space defined by pressure, semi-major axis,
obliquity, and eccentricity is not evenly sampled by our parameter
set; moreover, theoretically pressures, semi-major axes and initial
temperatures can take a range of values much larger than those we
used\footnote{We recall that initial temperatures are not part of the
  parameter space since they are initial conditions for each
  quadruplet of values in the parameter space itself.}.

For this reason, we define a volume element of the (five dimensional)
parameter and initial conditions space as follows:

\begin{equation}
\label{eq:dV}
dV[\vec{i},i_T] = dP(i_p) \cdot dR(i_r) \cdot dO(i_o) \cdot dE(i_e) \cdot dT(i_T)
\end{equation}

where $\vec{i}={i_p, i_r, i_o, i_e}$ is the position of a quadruplet
in the parameter space, and the differentials are the difference on
each of the five axis between two subsequent values of any parameters,
e.g. $dP(i_p) = P(i_p+1) - P(i_p)$. The integer indices $i_p, i_r,
i_o, i_e, i_T$ refer to the corresponding values in Table
\ref{table:dataset}. We extend the definition for the last point of
each set by taking the interval equal to the point before the last
one, e.g. $dP(i_{p{\rm max}}) = dP(i_{p{\rm max}}-1)$. Using Equation
\ref{eq:dV} the probability of SNOWBALL of one quadruplet becomes

\begin{equation}
\label{eq:prob}
P_{\rm snow} = \frac{\sum_{i_T=1}^{29} dV[\vec{i},i_T]_{\rm snow}}{\sum_{i_T=1}^{29}dV[\vec{i},i_T]}
\end{equation}

where $dV[\vec{i},i_T]_{\rm snow}$ are the volume elements
corresponding to cases of the quadruplet $\vec{i}$ that belong to the
class SNOWBALL while $dV[\vec{i},i_T]$ are all the volume elements of
{\it valid} cases of $q$. Of course, $P_{\rm warm} = 1-P_{\rm
  snow}$. Note that the probability defined in Equation \ref{eq:prob}
is automatically normalized to one, takes into account the uneven
sampling of the parameter space, and the fact that for some quadruplet
some of the runs in the initial temperature set could not have
converged.

Obviously, Equation \ref{eq:prob} can be generalized, e.g., to
estimate the overall probability to have a SNOWBALL or a WARM case,
just by summing over all the runs (dropping $q$ in the equation), or
by defining marginal probabilities and integrating on some of the
parameters.

The estimated probability $P_{\rm snow}$, as we defined it, clearly
depends on its support. The semi-major axis, the eccentricities and
the obliquities are evenly spanned, with the exception of the cases
$23.4393$ for the obliquity, $0.0167$ for the eccentricity, which we
included to have Earth conditions in the dataset. ESTM does not allow
us to study obliquities larger than 40 degrees, and we verified that
already at 40 degree the model shows unexpected behaviour. The same
holds for pressures: values lower than about 0.01 bar and higher than
10 bar are outside the range of validity of our model (see \citet{Vladilo15)}. 
Semi-major axes smaller than 0.9 AU or larger than 1.5 AU
can be studied with ESTM, but when the central star is Sun-like, they
are uninteresting as they always end respectively in snowball or
runaway greenhouse situations. For these reasons, the parameter space
is relatively well defined.

Instead, the initial temperatures are critical. While it is not
reasonable to start with very hot or very cold cases, that would
behave, in our ESTM model, as very small or very large distances from
the star, the specific definition of the probabilty support in the
intermediate initial temperature range do change the estimate of the
fraction of cases ending up in one or the other state.  \footnote{similar considerations 
can be drawn also for Daisyworld models in terms of the solar luminosity parameter
\citep{Adams, Alberti2015}.}. For this
reason, in the following we define three different supports, where we
keep always the whole set of pressures, semi-major axes, obliquities
and eccentricities but use:

\begin{itemize}
\item The whole initial temperature range of our dataset. Since we span a larger interval of initial temperatures reaching values lower than the freezing temperature, we call this support {\it  IceFavoured}
\item Temperatures from $i_T=-15$ to $i_T=20$, where $i_T$ is the index we used in Equation \ref{eq:tinit}. We call this support {\it IceCentered}
\item Only initial temperatures above the freezing point of water, from $i_T=0$ to $i_T=20$. This support is called {\it Liquid}.
\end{itemize}

In Figure \ref{fig:generalpies} we show the probability of having WARM
or SNOWBALL cases, for the three different probability supports
defined above, integrating over the whole parameter set and initial
conditions. The overall trend is as expected, with the {\it Liquid}
probability support giving more WARM cases than the {\it IceFavoured}
one, and the {\it IceCentered} support taking an intermediate
value. However, the fact that the figures are not dramatically
different suggests that our choice of the parameter space volume is
reasonable; for example, including more large semi-major axis values
would have simply increased the number of SNOWBALL situations.

To further explore these states, Figure \ref{fig:plotParamIce} shows
the marginal probabilities of ending in a SNOWBALL or a WARM state
depending on the value of each of the four parameters considered,
integrating over all values of the remaining three parameters and
initial conditions, when the IceCentered support is used. Similar
results are obtained for the Liquid support, in that case with a
larger fraction of WARM states.

From these overall results, several observations can be drawn:
\begin{itemize}
\item As expected, the probability of SNOWBALL increases for larger values of the SMA, independent of the other parameters. The SNOWBALL state becomes the only possible state for SMA about 1.5 AU.
\item The probability of WARM states grows with the pressure, integrating over all other parameters and initial conditions.
\item The marginal probabilities of WARM and SNOWBALL states remains approximately the same for eccentricity lower than about 0.3, then the probability of WARM states grows rapidly at larger values of the eccentricity. In some of such cases, the planet goes to a runaway greenhouse state.
\item Obliquity does not have a strong influence on the probabilities of WARM and SNOWBALL states, with a rather small trend towards more SNOWBALL states for larger obliquities.
\item In general, the initial temperature leading to SNOWBALL in the bistable cases is less than or equal to the freezing point of water (figure not shown). Higher initial temperatures lead to the WARM state.
\end{itemize}

\begin{figure*}
\centering
\includegraphics[width=0.60\textwidth]{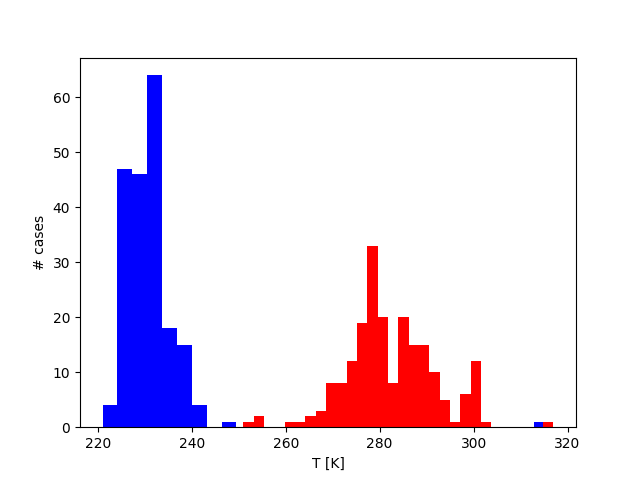}
\caption{Histogram of the final annual global average temperatures for the bistable quadruplets, in blue the histogram for the lower final annual global average temperatures, in red the istogram for the higher ones. Note that the histograms are made using the number of {\it cases} not that of {\it quadruplet} }
\label{fig:histT-bistab}
\end{figure*}

\begin{figure*}
\centering
\includegraphics[width=0.44\textwidth]{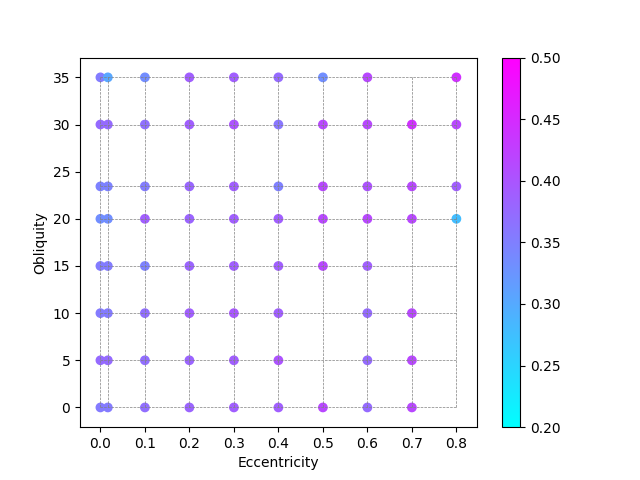}
\includegraphics[width=0.44\textwidth]{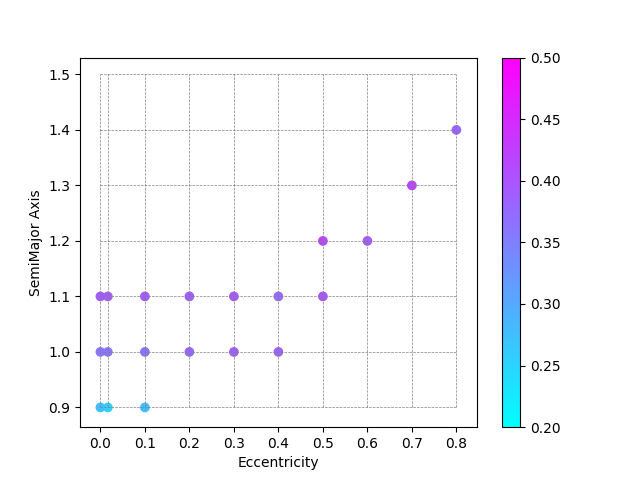}
\includegraphics[width=0.44\textwidth]{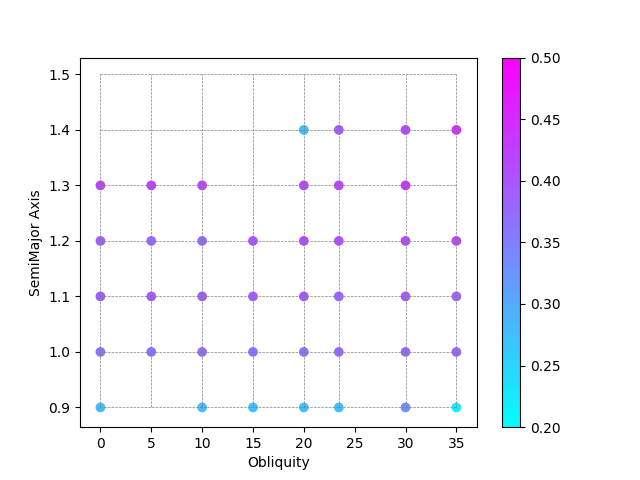}
\includegraphics[width=0.44\textwidth]{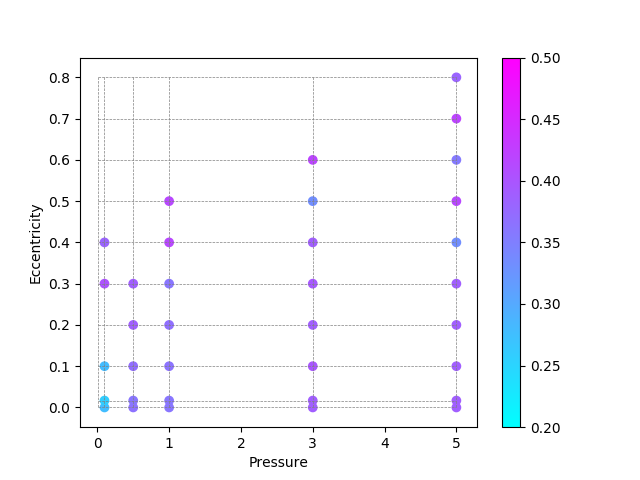}
\includegraphics[width=0.44\textwidth]{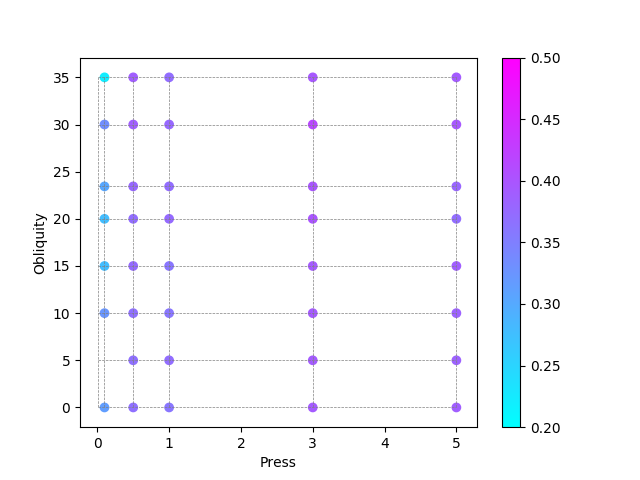}
\includegraphics[width=0.44\textwidth]{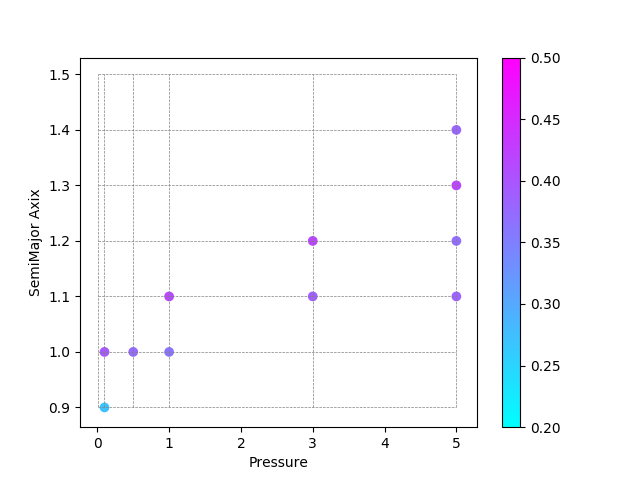}
\caption{
2D marginal probabilities from the 4D parameter space for the probability support {\it IceCentered}. In the panels we show the probability of having a SNOWBALL case as a function of all the couples of parameters, color coded as shown in sidebars. The probability is calculated using Equation \ref{sec:prob}, where the summation in Equation \ref{eq:dV} runs on the indexes that refers to the parameters not in the plotted couple. The grid shows the parameters we used in the dataset. We only plot dots where bistability happens; no dot at an intersection means that for that couple of parameter values there are no bistable cases. Upper left, upper right, middle left, middle right, lower left, lower right panes respectively to the couple of parameters:eccentricity \& obliquity, eccentricity \&semi-major axis, obliquity \& semi-major axis, pressure \&eccentricity, pressure \& obliquity, pressure \& semi-major axis.
}
\label{fig:bistab2Ice}
\end{figure*}

\begin{figure*}
\centering
\includegraphics[width=0.60\textwidth]{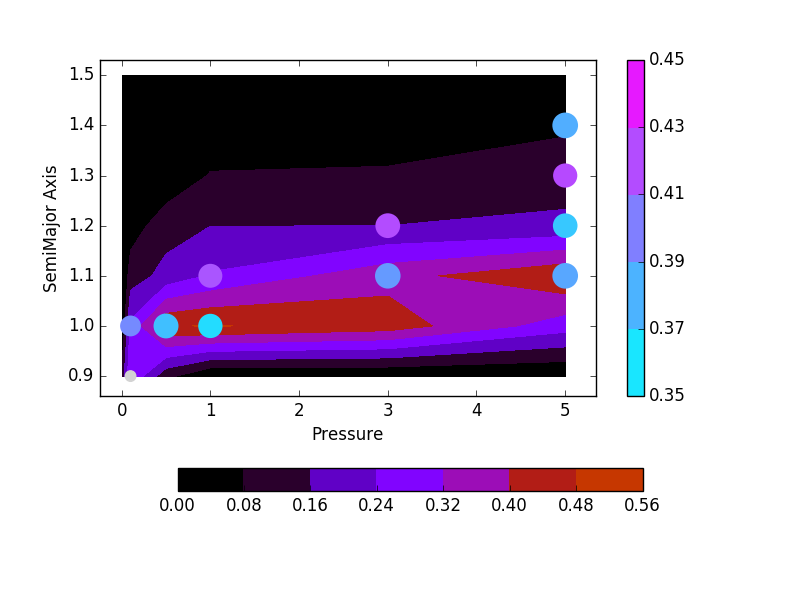}
\includegraphics[width=0.60\textwidth]{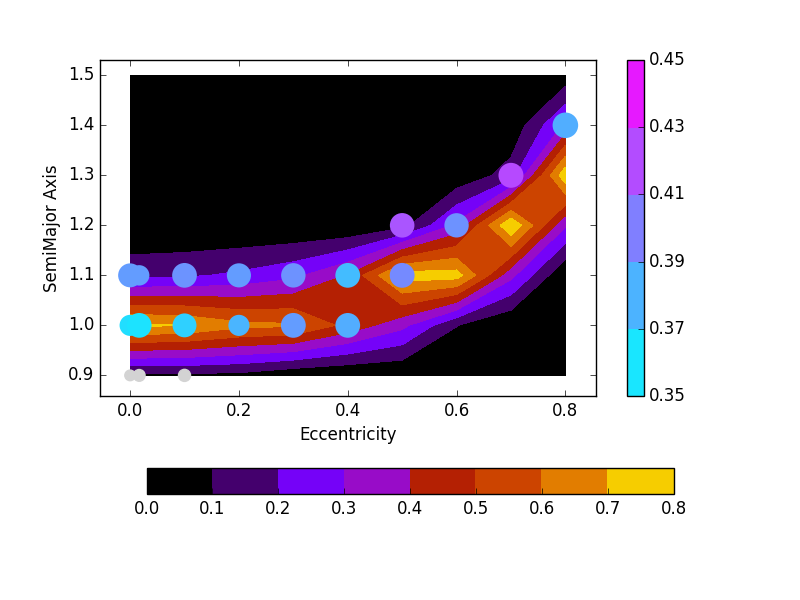}
\caption{
2D marginal probability from the 4D parameter space for the probability support {\it IceCentered}, comparing bistability and habitability. The points indicate bistable solutions and the color coding of the points indicates the probability of having a SNOWBALL case, color coded as in the vertical bar. Small grey points indicate bistable cases with a SNOWBALL probability lesser than 0.35; they are rare and usually get values of about 20-25\%. The color map indicates the habitability parameter (between zero and one) for complex life, $h_{050}$, as color coded in the horizontal bar and obtained using the definition from Silva et al. (2016). Top panel: Pressure {\it vs.} Semi-Major Axis. Bottom panel: Eccentricity {\it vs.} Semi-Major Axis.
}
\label{fig:bistab-hab} 
\end{figure*}

\section{Bistability of the Earth case in ESTM}
\label{sec:earth}

Using the ESTM, we first investigated the possible presence of
multiple equilibrium states for a set of conditions which are typical
of the Earth. These are summarized by the following quadruplet:
Pressure $=1.0$ bar, Semi-Major Axis $=1.0$ AU, Obliquity
$=23.4393^\circ$, and Eccentricity $=0.01671022$.  Thus, for the same
value of the solar constant, we chose several different values of the
initial temperature (as detailed in Table \ref{table:dataset} ) and we
calculated the annual mean global temperature of the Earth reached in
the stationary state after the initial transient. The results are
shown in Figure \ref{fig:earth}: two main climatic regimes are
identified, representative of the SNOWBALL and WARM regimes,
characterized by a temperature range of 25.90 K - 260.66 K and 265.66
K - 370.55 K, respectively. Thus, the transition between one state and
the other occurs for an initial temperature value between 260 and 265
K.  Therefore, for the present-day value of the solar constant, in a
given range of initial temperatures the global mean temperature
converges to 289.42 K, corresponding to the warm case, while for
another range of initial temperatures the solution converges to 231.79
K, corresponding to the fully ice-covered case. We can summarize these
findings as follows:

\begin{itemize}
  
\item The current astronomical conditions of the Earth support two
  possible steady climate states, one corresponding to the present
  climate and the other characterized by much lower
  temperatures. Depending on the initial temperature, the system is
  attracted to one of these states (see also \citet{Alberti2015}
    and \citet{Rombouts2015}) .

\item Under current initial temperature conditions (in the range of
  liquid water), the transition to a snowball state is unlikely,
  unless a change in insolation or in other elements of the Earth's
  radiation budget, such as one caused by strongly reduced levels of
  CO$_2$ or other Green-House gases (GHG)s, or by a strong increase in
  stratospheric aerosols, occurs.

\end{itemize}

An important point is that we verified that the bistability observed
here is driven by the ice-albedo feedback. In fact, we run another
``Earth'' quadruplet artificially removing the ice-albedo feedback: in
this case, no bistability develops. Furthermore, both for a smooth
transition between bare surface and ice cover (as adopted in the
standard ESTM) and for ``istantaneus'' ice - no ice transition (that
is, ice appears as soon as the temperature drops below the freezing
point of water), we recover the bistability. Noticeably, the
transition between the two states happens at the same limit
temperature. This suggests that the behaviour of the climate in ESTM
does not depend on the detailed parametrization of ice but only on its
presence and response to temperature.

These results show that the ESTM provides results on Earth climate
bistability which are consistent with those found using simpler Energy
Balance Models. For example, also in studies using EBMs, the
likelihood to enter a snowball-Earth climate seems to be quite small
\citep[e.g. ][]{Zalia10}.

\section{Climate bistability in earth-like exoplanets and its links with habitability}
\label{sec:prob}

To proceed with a full exploration of the climate bistability in earth-like
exoplanets, we first need to properly define: (i) what we mean with
{\it stability} and {\it bistability}; (ii) how we define and compute
probabilities.

We define a {\it stable} solution, one for which all the initial
temperatures for a given quadruplet of parameter values produce {\it
  the same} final annual global average temperature. In dynamical
systems terms, this is a {\it globally stable} solution. A {\it
  bistable} solution instead occurs when two of such final
temperatures are produced for the same quadruplet, depending on the
initial conditions. Multistable solutions are characterized by a
larger number of stable states, but these solutions have not been
found in our exploration of the ESTM.

Note that we did detect quadruplets showing more than two
temperatures, as described in Section \ref{sec:dataset}. However, we
verified that in none of the such quadruplets, we had a situation with
convergence to more than two final stable states with different
temperatures. Rather, is such cases oscillating behaviours and/or no
convergence after 100 orbits was observed. All those cases are
peculiar and genuine tri-stability never occured with ESTM, at least
within the parameter space which we explored\footnote{For three
  quadruplets, 28 initial temperatures converged to a low or a high
  final value, and only one did not converge after 100 orbits. We
  discarded these quadruplets.}.

The following question is how many quadruplets are associated with two
possible final states, reached for different values of the initial
conditions. These are the true bistable cases. We identified such
bistable cases by searching for the quadruplets where the final
average temperatures display significant differences depending on the
initial temperature. This search provided 179 quadruplets that are
associated with bistable solutions. Figure \ref{fig:histT-bistab}
shows the histogram of the higher (red) and lower (blue) final
temperature when we had bistability (compare it with Figure
\ref{fig:histT-all}). Note, also, that when a quadruplet shows
bistability, almost always the runs ending in the lower temperature
state are SNOWBALL and the others are WARM. All the other quadruplets
lead to a single final stable solution.

For the bistable quadruplets, Figure \ref{fig:bistab2Ice} shows 2D
projections (i.e., marginal probabilities) of the 4D parameter space
for the probability support {\it IceCentered}. In the different panels
we show the marginal probability of having a SNOWBALL case as a
function of a given parameter couple, color coded as shown in sidebars
and integrated over the other two parameters. Some of the quadruplets
have almost equal probability of ending up in a WARM or SNOWBALL state
depending on initial temperature.

From this figure, a few comments can be drawn:

\begin{itemize}
\item {Bistable cases are found for almost all parameter values of pressure, eccentricity and obliquity.} 
\item Pressure and obliquity have rather limited effect on the existence of bistability, that is, one can find bistable solutions for any value of these parameters;
\item bistability is confined to a well defined range for the semi-major axis, a behavior which is especially visible in the SMA-pressure and SMA-eccentricity plots. 
\end{itemize}

Clearly, the quantitative aspects of these results depend on the
support chosen for the initial temperatures: using only initial
temperatures above the freezing point drastically reduces the
probability of ending up in a SNOWBALL state and thus the presence of
bistability.

A final important point concerns the relationships between bistability
and habitability. From Figure \ref{fig:bistab2Ice}, it is clear that
bistability is confined to a rather narrow range of orbital semi-major
axis values, which varies with eccentricity and pressure. Figure
\ref{fig:bistab-hab} compares the range where bistability was observed
with the range where the conditions for the presence of complex life,
as defined by \citet{Laura16}, are fulfilled. Top panel shows
  the connection among pressure, semi-major axis, bistability and
  habitability. Bottom panel show the same, but using eccentricity.

This figure reveals the intriguing result of a close similarity
between the ranges where ``complex life habitability'' and bistability
are found: if a planetary climate fulfills the habitability criterion,
then the same climate is potentially bistable. The interpretation of
this result in terms of thermal dependence of biological processes is
beyond the purpose of the present work. Here we note that the
conditions of bistability allow for the simultaneous presence of two
fundamental feedbacks of the climate systems, namely the ice-albedo
feedback and the temperature-water vapour feedback. Since these two
feedbacks have opposite responses to variations of insolation, they
may help to provide a long-term persistence of liquid-water
habitability conditions.

We also note that the vast majority of bistable {\it and} habitable
cases have similar probabilities of being snowballs or warm.

In Figure \ref{fig:bistab-hab}, we note that at high eccentricities
and relatively low values of the semi-major axis, high habitability
does not correspond to bistable cases. This is because, at these
extreme eccentricities, the planet can be warmed-up very fast at the
pericenter; this melts the initial ices. Then, only if the semi-major
axis is large enough, ice can re-form and bring the system again to a
SNOWBALL case. This behaviour {\it does} depend on the details of the
ice treatment that in ESTM is, of course, simplified. In these extreme
cases, a study with intermediate-complexity models, e.g. PLASIM
\citep{Fraedrich2012}, would be needed.

\section{Discussion and Conclusions}
\label{sec:conclu}

The main results of this work are that
\begin{itemize}
\item Climate bistability is relatively common but not widespread in planetary climates. For 179 quadruplets over the total of 3051 valid quadruplets explored in our study, bistability was clearly detected. In no case we observed more than two stable states. For all other quadruplets for which converge occurred, the climate ended in either a WARM or a SNOWBALL state. 
\item Bistability generally includes a WARM state, similar to the current state of our planet, and a SNOWBALL state, presumably analogous to what happened to Earth in its past history.
\item Although not widespread, the parameter range where bistability occurs is quite similar to the range where habitability conditions for complex life are present.
\end{itemize}

These results have a few implications that are worth
discussing. First, the model runs suggest that climate bistability can
occurr in earth-like exoplanets under a variety of parameter values, as
suggested by \citet{Kilic17} for aquaplanets. Thus, for the same
external control parameters (stellar luminosity, orbital
characteristics and gross atmospheric properties), the planet can
either be in a WARM state or in SNOWBALL conditions. Clearly, the two
states lead to very different environmental and planetary conditions.

We verified that in our model the transition to a snowball state
usually happens when the initial temperature is below the freezing
point of water. In our ARTECS database, we always used a ``cold
start'' initial condition, but here ``cold'' means ``sligthy above the
freezing point''. The current work thus suggests that the database
initial conditions are appropriate to maximize the WARM cases. In an
ongoing work, we will use ARTECS to better characterize the
habitability of earth-like exoplanets, keeping in mind that we are using
the most favourable temperature initial condition for complex life.

In our simple modelling approach, the planet can end up in either a
SNOWBALL or WARM state depending on initial conditions, without any
possibility to alternate between the two states (as happened to
Earth). In reality, a variety of mechanisms can trigger the transition
from a WARM to a SNOWBALL state or viceversa. One essential mechanism
is the reduction (for triggering the SNOWBALL) or accumulation (to end
the SNOWBALL period) of carbon dioxide in the atmosphere, owing to
volcanic outgassing, the geological carbon cycle, extreme weathering
or biospheric activities \citep[see e.g. ][]{Broecker18}. In our model, the
carbon dioxide content of the atmosphere is fixed and no transitions
can be observed. Future work will consider an imposed variable
atmospheric carbon dioxide concentration as well as the inclusion of a
simplified carbon cycle in the ESTM.

Finally, one of the most intriguing results of this work is the
indication that the range of values where climate bistability was
found is very similar to the range where the planet can host complex
life. This certainly occurred for Earth, and one can wonder whether
there may be deeper links between the presence and evolution of
complex life and the possibility of entering and exiting a SNOWBALL
state \citep{RareEarth}, or even between the presence of climate
bistability and plate tectonics activity through its effects on the
geological cycle of carbon dioxide. Clearly, these issues cannot be
addressed with the simplified approach adopted here, but our results
point to the need for further exploring these aspects. Finally, we
note that, at least for the Earth case, in the model the transition
occurs at an {\it annual global average} temperature of -13$^\circ$
C. This means that strong changes in the climate forcings are required
to initiate a snowball.

\section*{Acknowledgements}

Simulations were partly carried out using time granted under the CHIPP INAF initiative, at the Trieste local intel cluster. Part of the post-processing has been performed using the PICO HPC cluster at CINECA through our expression of interest. G.M. acknowledges the Geoscience and Earth Resources Insitute of National Research Council of Italy (CNR) for financial support. We thanks the anonymous referee for useful suggestions. The Authors wish to thank the Italian Space Agency for co-funding the Life in Space project (ASI N. 2019-3-U.0)


\bibliographystyle{mnras}



\appendix

\section{ARTECS}

Our ARTECS (Archive of terrestrial-type climate simulations) database is aimed at parameter space exploration of exo-climes using ESTM. Currently, we investigated changes in pressure, semi-major axis, eccentricity, obliquity,  planet geography and CO$_2$ partial pressure. Parameter values can be found in Table \ref{table:artecs}. All runs have ``cold start'' initial conditions, thus a constant temperature $T=275$K at all latitudes.

ARTECS is public and can be reached at the site {\rm http://wwwuser.oats.inaf.it/exobio/climates/}.
The user can make queries to the archive using parameter values, results (e.g. surface temperature, habitability, etc.), or both.

Results of queries can be downloaded in form of text tables, if only a summary is needed. Detailes FITS maps of surface temperatures {\it vs.} time (i.e., orbital position) and latitude can also be downloaded.

ARTECS currently contains 20284 cases: we put on the archive only ``converged'' runs, thus automatic snowballs, runaway greenhouse cases and failed integrations are not reported.
We plan to extend the archive to also varying planet radius, rotation period and surface gravity.
To allow scriptin access to ARTECS, a Python3 interface to access
ARTECS (py\_artecs) have been prepared by our group. The interface with
instructions to download, install and use it can be recovered from the
ARTECS website or through git at the url
https://www.ict.inaf.it/gitlab/michele.maris/py\_artecs.git

\begin{table*}
	\centering
	\caption{ Parameters varied in the ARTECS archive. Pressures in bar; Semi-Major axis in AU; Obliquities in degrees, CO$_2$ partial pressure in Earth units (we used 380 ppm). Ocean fraction only applies to the first geography.}
	\label{table:artecs}
	\begin{tabular}{lcl} 
		\hline
		Parameters & Value \\
		\hline
		Pressures & 0.01, 0.1, 0.5, 1.0, 3.0, 5.0 \\
                Semi-Major Axis & 0.9, 1.0, 1.1, 1.2, 1.3, 1.4, 1.5 & \\
		Obliquities & 0., 5., 10., 15., 20., 23.4393, 30., 35., \\
		Eccentricities &  0.0, 0.01671022, 0.1, 0.2, 0.3, 0.4, 0.5, 0.6, 0.7, 0.8 & \\
                Geography & constant fraction of oceans, equatorial continent, polar continent, current earth \\
                Ocean fraction & 0.1, 0.2, 0.3, 0.4, 0.5, 0.6, 0.7, 0.8, 0.9 \\
                CO$_2$ partial pressure & 0.1, 1, 10, 100\\
		\hline
	\end{tabular}
\end{table*}

From the technical point of view,
implementing a strategy for the preservation, curation and access to data is of vital importance in order to guarantee data reusability and reproducibility of scientific results.
To this end, an archiving software, a database and an interface for data retrieval are needed.
Developing an archiving software may be very challenging especially because of data evolution.
The main issues comprise changes in data format, variations in publication policy and in metadata content.
For all these reasons a good archiving software needs to be flexible and configurable, reusable in different contexts, easily deployable and scalable.
The New Archiving Distributed InfrastructuRe (NADIR) developed within the IA2 (Italian Astronomical Archive) project in takes all these requirements into account. It is based on TAco Next Generation Object (TANGO, www.tango-controls.org) and it inherits all its points of strength, like standardization of logging, high scalability, modularity and robustness.
Each TANGO device is an archiving module performing a specific task, for example sorting out a certain data format, checking its consistency with the standard, writing the metadata to a MySQL database, transferring data and metadata over the Internet to the archiving sites or receiving the data from the archiving stations.
Once a device has been developed, it can quickly be deployed on different archiving machines and its properties are modifiable at any time through a graphical interface.
The flexibility deriving from these conditions sets the basis for a robust and reliable service.
Furthermore, TANGO distributed control system is responsible for service start up, shut down and keep alive processes.

For the realization of the exoclimates archive, two specific TANGO archiving devices are used: \textit{preProcessor} and \textit{fitsImporter}.
The files resulting from the simulations are copied to an incoming folder on which the \textit{preProcessor} device is active.
In particular, it performs a selection of the correct file format and checks the conformance of the file to the FITS standard.
In case of a positive outcome, the file is transferred to a second folder, subject to the action of the \textit{fitsImporter}.
This second device extracts the metadata from the FITS cards of the selected files according to the keys stored in the datamodel and fills the instrument tables of a MySQL database.
It finally moves the file to its permanent position in the storage.

Database generated by NADIR can be queried using a TAP service. TAP stands for Table Access Protocol and is a Virtual Observatory (VO) standard which allows performing custom queries on astronomical databases using ADQL, an astronomy-oriented SQL-like language. The exoclimate archive web interface is built on the top of the TAP service using APOGEO (Automatic POrtal GEneratOr), a wizard for the creation of astronomical web portals which has been used to generate all the most recent IA2 portals.
Using the portal, the users can apply various filters to their searches on the archive and download the files included into the search result. Moreover it is possible to select multiple files and download a tar archive of all of them or export the search result in CSV or VOTable format. A VOTable is a XML file format standardized by VO in order to store tabular data in an interoperable way.
VOTables and FITS files can also be sent to external VO applications using SAMP (Simple Application Messaging Protocol), another VO standard for exchanging astronomical data between VO-compliant applications. A typical use case can be sending a VOTable to TOPCAT, an application for performing operation on catalogues and tables.
It is also possible to download a list of all the files resulting from a search. This can be useful if one wants to download a set of files using a non-browser client (e.g. wget).
Internally, the web portal is composed by multiple web services working together: the TAP service mentioned before, a File Server (for retrieving FITS files and their preview images), an User Space service (for storing user generated temporary files, like VOTables or tar files) and the web application containing the search form. All these applications are written in Java EE and can be deployed both on GlassFish or Tomcat servers.

\bsp	
\label{lastpage}
\end{document}